\documentclass[sigplan]{acmart}
\usepackage{amsmath,amsfonts}
\usepackage{algorithmic}
\usepackage{graphicx}
\usepackage{textcomp}
\usepackage{xcolor}

\usepackage{tabularx}
\usepackage{supertabular}

\usepackage{amsthm}
\usepackage{amsmath}
\usepackage[]{caption}
\usepackage[skip=3pt]{subcaption}
\usepackage{multirow}
\usepackage{mathrsfs}
\usepackage{amsbsy}
\usepackage[mathscr]{eucal} %% For \mathscr

\usepackage{comment}
\usepackage{pifont}% http://ctan.org/pkg/pifont

\AtBeginDocument{%
  \providecommand\BibTeX{{%
    \normalfont B\kern-0.5em{\scshape i\kern-0.25em b}\kern-0.8em\TeX}}}

\copyrightyear{2023}
\acmYear{2023}
\setcopyright{rightsretained}
\acmConference[e-Energy '23 Companion]{The 14th ACM International
Conference on Future Energy Systems}{June 20--23, 2023}{Orlando, FL, USA}
\acmBooktitle{The 14th ACM International Conference on Future Energy Systems
(e-Energy '23 Companion), June 20--23, 2023, Orlando, FL, USA}

\acmDOI{10.1145/3599733.3600257}
\acmISBN{979-8-4007-0227-3/23/06}

\begin{document}

\title{Power Grid Behavioral Patterns and Risks of Generalization in Applied Machine Learning}

\author{Shimiao Li}
\email{shimiaol@andrew.cmu.edu}
%\email{webmaster@marysville-ohio.com}
\affiliation{%
  \institution{Carnegie Mellon University}
  %\streetaddress{P.O. Box 1212}
  %\city{Pittsburgh}
  %\state{PA}
  \country{USA}
}

\author{Jan Drgona}
\email{jan.drgona@pnnl.gov}
\author{Shrirang Abhyankar}
\email{shrirang.abhyankar@pnnl.gov}
\affiliation{%
  \institution{Pacific Northwest National Laboratory}
  %\city{Pittsburgh}
  %\state{PA}
  \country{USA}}

\author{Larry Pileggi}
\email{pileggi@andrew.cmu.edu}
\affiliation{%
  \institution{Carnegie Mellon University}
  %\streetaddress{P.O. Box 1212}
  %\city{Pittsburgh}
  %\state{PA}
  \country{USA}
}


\begin{abstract}
  Recent years have seen a rich literature of data-driven approaches designed for power grid applications. However, insufficient consideration of domain knowledge can impose a high risk to the practicality of the methods. Specifically, ignoring the grid-specific spatiotemporal patterns (in load, generation, and topology, etc.) can lead to outputting infeasible, unrealizable, or completely meaningless predictions on new inputs. 
%Meanwhile, the limited public access to real-world data can prevent us from understanding the grid-specific patterns. This raises the fear of training and evaluating a model on synthetic data that are generated unrealistically. 
To address this concern, this paper investigates real-world operational data to provide insights into power grid behavioral patterns, including the time-varying topology, load, and generation, as well as the spatial differences (in peak hours, diverse styles) between individual loads and generations. Then based on these observations, we evaluate the generalization risks in some existing ML works caused by ignoring these grid-specific patterns in model design and training.  
%The findings in this paper will serve to facilitate the incorporation of domain knowledge in the context of power grid, towards improved performance of data-driven methods for practical use.

\end{abstract}



\keywords{domain knowledge, generalization, machine learning, power grid}


\maketitle

\section{Introduction}
\label{sec:Introduction}

Recent years have seen a rich literature of data-driven approaches designed for the power grid. A wide range of traditional and modern threats \cite{grid-cybersecurity-Vyas}, ranging from natural faults to targeted attacks \cite{fdia-review} \cite{ukrain2015} \cite{MadIoTOnNY}, have motivated the use of anomaly detection approaches \cite{dynwatch} to safeguard the power grid. The computation burden of solving nonlinear optimization problems in operation and planning has motivated the development of data-driven alternatives to state estimation (SE) \cite{gnn-se} \cite{rnn-dse}, power flow (PF) analysis \cite{pf-dnn-topo}\cite{physics-NN-pf}, optimal power flow (OPF) \cite{dc3} \cite{homotopy-pnnl}\cite{data-driven-optimization-survey}, as well as data-driven warm starters to collaborate with physical solvers \cite{gridwarm}\cite{dnn-opf-warm-starter}, etc. 
%However, an insufficient consideration of domain knowledge can impose a high risk to the practicality of data-driven methods. This includes but is not limited to the follows: methods based on supervised learning (like some neural network models) require training on labeled data, which are expensive to obtain (e.g., labelled OPF data needs a lot of computation) or even unavailable (anomaly data are typically unlabelled) on a realistic grid; the performance degrades when facing high-dimensional data from a large-scale system; the correlations learned from data-driven models can be inaccurate when compared with the real physical constraints; infeasible xxx. xxx

Despite their popularity in recent years,
%the use of general ML tools on power grids have long been found risky under realistic grid conditions.
people have long been aware of the risks of machine learning (ML) tools regarding their impracticality\cite{gridwarm_old} under realistic power grid conditions. The risks come from the "missing of physics" in general ML methods.
Specifically, 
%the temporal grid evolution is physically driven by transient system dynamics, changing topology, varying demand and re-dispatched supply, whereas the spatial correlations are enforced by network constraints, operational bounds and technical limits. 
the transient system dynamics, changing topology, and varying supply and demand
are physical reasons behind the temporal grid evolution. In contrast, the physical power flow constraints, operational bounds, and technical limits determine the spatial correlations between different locations.
To address the "missing physics", recent works insert domain knowledge into their models. Attempts range from creating domain-knowledge-based features \cite{hooi2018gridwatch} and inserting domain knowledge (e.g., adding topology information  \cite{gnn-se}, steady states~\cite{King2022}, or network constraints \cite{dc3} \cite{homotopy-pnnl}) into the models, to building hybrid approaches that combine machine learning with numerical solvers\cite{dnn-opf-warm-starter}.

To enhance the use of domain knowledge and develop practical models for grid-specific tasks, it is important to gain a clearer understanding of the realistic power grid behaviors, including both temporal and spatial patterns. Yet, limited public access to real-world data can prevent us from gaining insights. This raises the fear of insufficiently considering the power grid reality and training a model on synthetic data that are generated unrealistically. To address these concerns, this paper investigates real-world power grid data to visualize the spatiotemporal behavioral patterns of the following:
\begin{itemize}
    \item temporal evolution of network topology, total supply and demand
    \item temporal pattern and spatial differences (including different peak hours, different variation styles, etc.) in individual loads and generations
\end{itemize}

Unless we have a huge (nearly infinite) amount of data to define all possible scenarios, ignoring these grid-specific behavioral patterns can cause generalization risks, making ML models producing inaccurate, infeasible and even meaningless outcomes.
In this paper, we evaluate %how the non-consideration of changing topology and spatial differences in load variations can degrade the performance in power grid anomaly detection and data-driven optimal power flow. Our results demonstrate 
the following generalization issues:

\begin{itemize}
    \item ML models ignoring system topology cannot generalize on dynamic graphs: we evaluate this risk by anomaly detection scores in the context of topology change
    \item ML models trained without consideration of spatial differences in load/generation cannot generalize well on realistic system configurations: we evaluate this risk from anomaly detection scores, as well as solution feasibility of data-driven OPF models
\end{itemize}

These lessons of generalization risks motivate the consideration of realistic behavioral patterns in the design, and training of data-driven models for power grid applications.

\section{Spatiotemporal Behavioral Patterns and Constraints}\label{sec: grid conditions}

The power grid has a graphic nature.
The nodes which correspond to substation buses and transmission towers are inter-connected by edges that correspond to branches (transmission lines, transformers) and closed switch links. 
Among the buses, some are \textit{injection buses} with active generators and/or loads, leading to power supplied or consumed at these locations; others are \textit{zero-injection (ZI) buses} that have no generators or loads, making the sum of current or power over all associated branches for each ZI bus zero. 
The connectivity of these buses is identified as the network \textit{topology}. 

%The power grid's complicated design, functionality and control scheme of make it a dynamic graph with unique spatiotemporal behavioral patterns. 
%The power grid has a time-varying nature. The temporal evolution can be driven by system dynamics, changing topology, fluctuating demand and re-dispatched supplies. Whereas spatially, different buses can affect each other. The spatial correlations and inter-dependencies are enforced by network constraints, operational bounds and technical limits. But spatial differences also exist, a power grid change tends to most affect its local neighborhood most, and the impact declines as the distance increases.
%Understanding the power grid's unique spatiotemporal patterns is important for designing related practical methods. 
This section provides insights into the behavioral patterns of topology, load and generation on real-world large-scale power grids.  We analyze the underlying physical reasons behind these patterns to better understand  the power system. 

All observations are obtained from real data:

\textbf{Data description:} 24-hour data of system model description from a real utility in the Eastern Interconnect of the U.S. The dataset contains a full bus-branch model of the grid based on a 0.5-hour interval.   

\subsection{Network Topology Patterns}
\label{sec: patterns topology}

We first investigate how the power grid topology
behaves and evolves over time. Two types of changes are observed from the real data described above to reflect topology changes:

\begin{itemize}
    \item Change in active bus set, characterized by bus split and merge. In the control room, the network topology processor merges any two nodes that are connected by a closed switch. Thus, upon any switch closure, there will be two separate buses  merging into one, denoted as a \textit{bus-merge} in this paper.  Any switch opening may cause the previously merged buses to split, denoted as a \textit{bus-split}. 
    \item Change in branch status, characterized by the opening (disconnection) and closure (connection) of transmission lines and transformers. 
\end{itemize} 

\begin{figure}
     \centering
     \begin{subfigure}[t]{0.48\textwidth}
         \centering
    \includegraphics[width=0.8\textwidth]{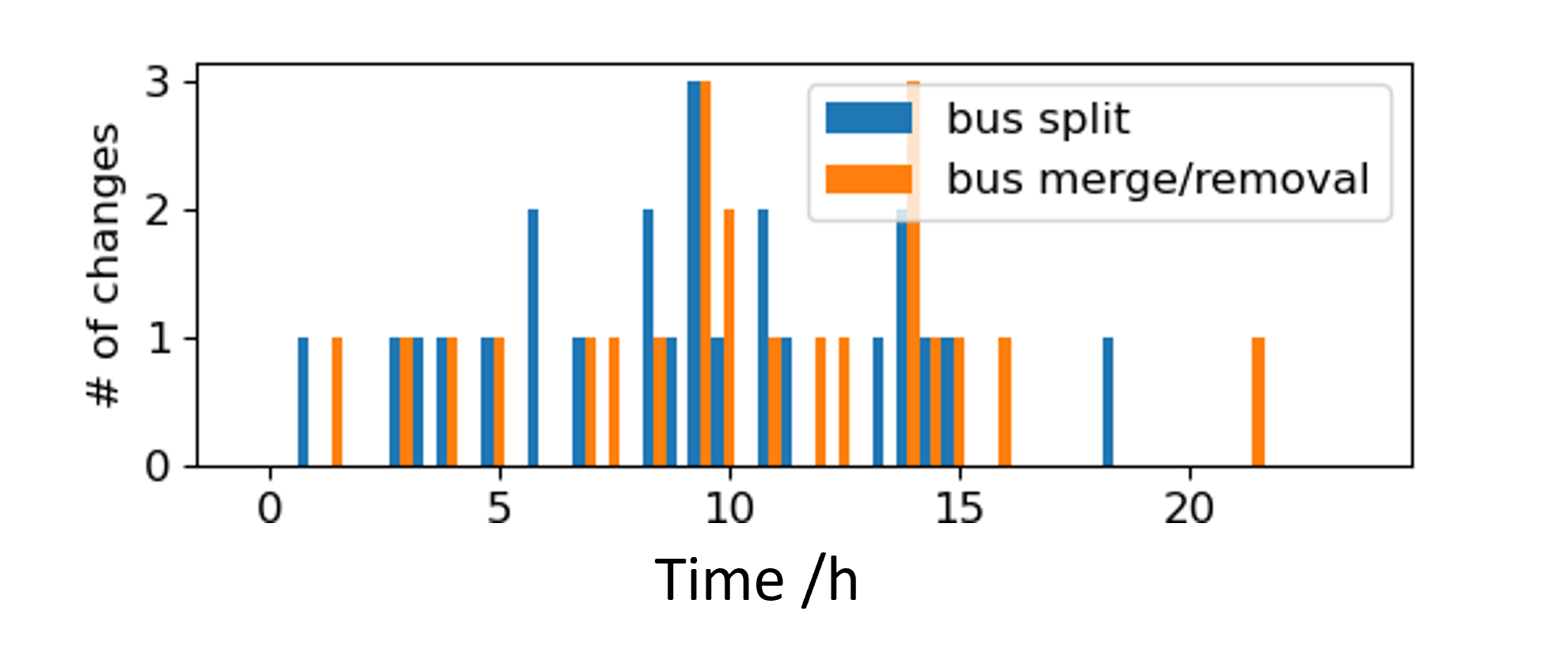}
         \caption{Record of bus split and merge: 46 operations observed in 24 hours, among approximately 20,000 buses.}
         \label{fig: topology bus}
     \end{subfigure}
     \hfill
     \begin{subfigure}[t]{0.48\textwidth}
         \centering
    \includegraphics[width=0.8\textwidth]{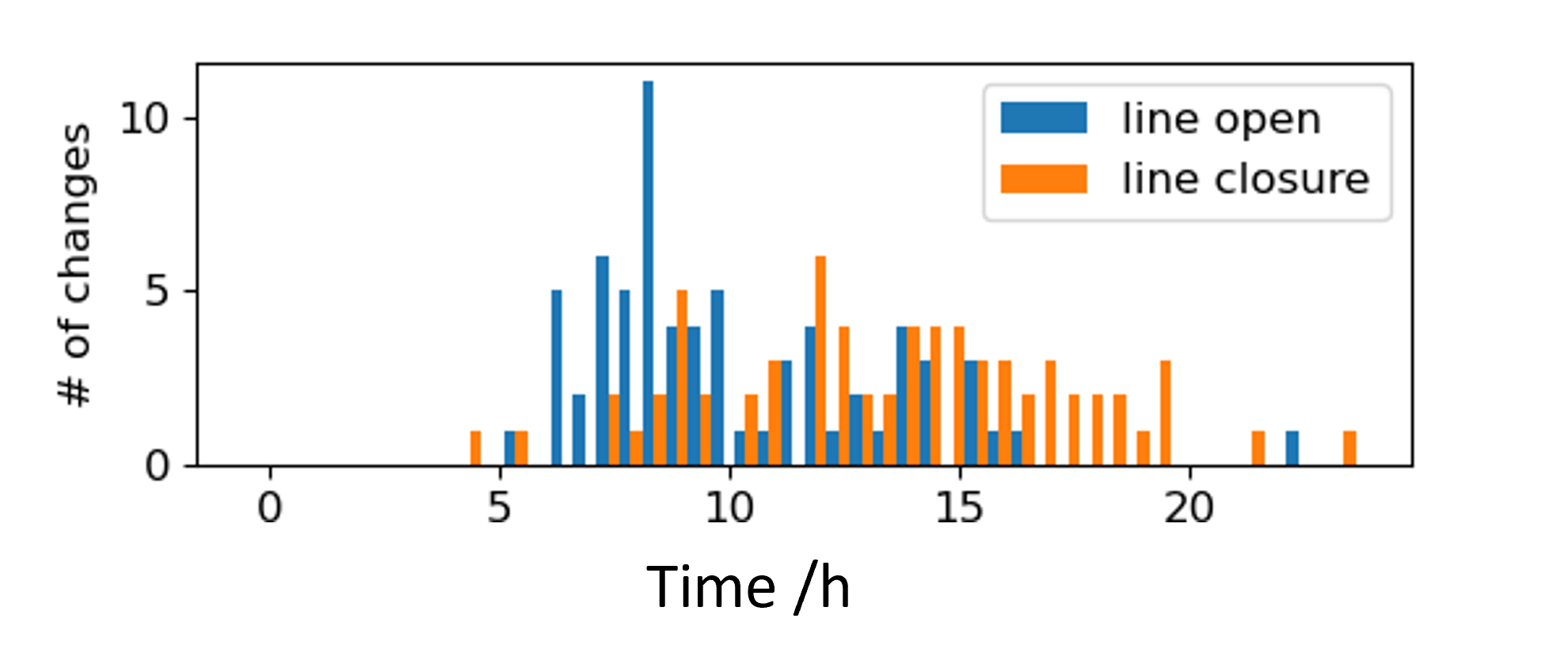}
         \caption{Record of transmission line status change: 137 status changes observed in 24 hours, among approximately 17,000 lines.}
         \label{fig: topology line}
     \end{subfigure}
     \caption{Topology changes in 24 hours.}
        \label{fig: topology}
\end{figure}
\textbf{Observation:} The (bus-branch) topology changes throughout the day. In Figure \ref{fig: topology bus}, 46 bus-splits and bus-merges are observed in a day on a grid with approximately 20,000 buses in total. In Figure \ref{fig: topology line}, 137 transmission line status changes are observed in a day on a grid with approximately 17,000 lines. 

\textbf{Physical reasons:} The change in topology can be either a control action or an unexpected event (anomaly). Examples of control actions include opening a transmission line or transformer for scheduled maintenance, reclosing a transmission line after tripping (transmission operation in PJM (a regional transmission organization that coordinates the movement of wholesale electricity in the United States) requires any extra high voltage (EHV) transmission line to be manually reclosed within 5 minutes after tripping, if it does not automatically reclose\cite{pjm-m03}),  
 opening transmission line for high voltage control \cite{pjm-m03},
closing a transmission line to meet the required transfer capacity, etc. Unexpected events can include sudden and unplanned line tripping (due to overloading, physical damage, etc.), transformer failures, circuit breaker failures, etc.

\subsection{Total Generation and Load Pattern}
\label{sec: patterns total load gen}

Apart from a dynamic topology, the supply and demand on the power grid are also subject to variation. This section investigates the behavior of total load and generation.

\begin{figure}[h]
\centering
\includegraphics[width=0.9\linewidth]{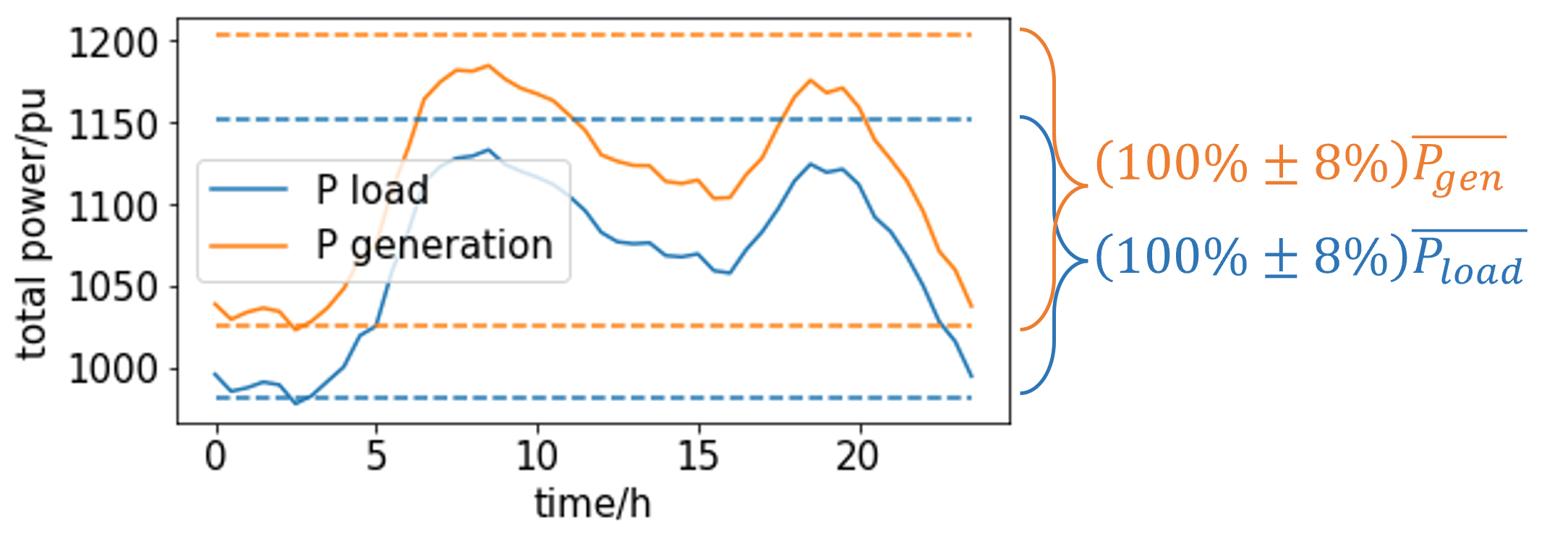}
\caption{Variation of total load and generation in 24 hours.}
\label{fig: total load and gen}    
\end{figure}

\textbf{Observation:} 
The total demand and supply on a large grid tend to peak in the morning and early evening. Figure \ref{fig: total load and gen} shows that the total daily load and generation vary smoothly within the range of $(100\%\pm 8\%)$ of the mean value, with the generation following the demand throughout the day.

\textbf{Physical reasons:} 
The total load demand depends highly on the weather conditions. Peak load periods tend to occur in the morning during winter months (when mass heating happens) and in the afternoon during summer months (mass cooling).
The major reason for the generation following the demand is the economic operation and optimal control of the power system. Approximately every 5-15 minutes (every 5 minutes in PJM \cite{pjm-m11}), the operators perform economic dispatch (ED) and security-constrained economic dispatch (SCED) to make the forecasted demand at the lowest possible generation cost (subject to security constraints, if in SCED). The ED solution primarily depends on the generating unit cost function. Approximately every 5 minutes to 1 hour, the operator performs optimal power flow (OPF) to determine the generators' optimal operating point (power output and voltage set point) that meets the demand with minimal loss or operating cost. Unlike economic dispatch, which is mainly for market purposes, OPF considers power flow constraints of the network and a variety of technical limits.

\subsection{Individual Load Patterns}
\label{sec: patterns load}

Despite the total load pattern observed, every individual load does not follow the same pattern as above. This section investigates the individual load patterns from the perspective of the peak load time, and the style of variation. 

\begin{figure}
     \centering
     \begin{subfigure}[t]{0.3\textwidth}
         \centering
    \includegraphics[width=\textwidth]{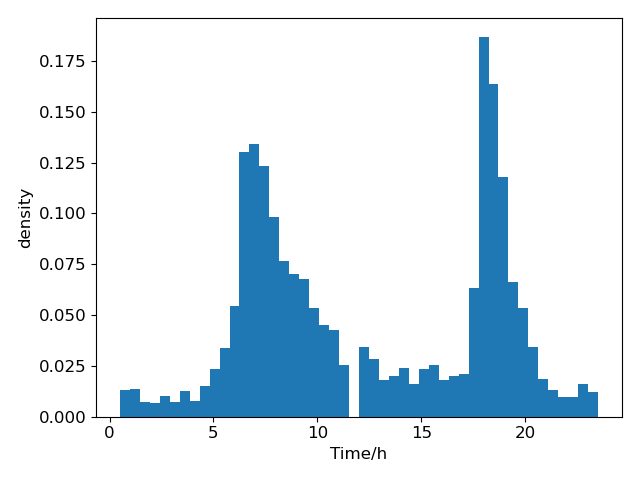}
         \caption{Distribution of peak load time.}
         \label{fig: load peaktime dist}
     \end{subfigure}
     \hfill
     \begin{subfigure}[t]{0.48\textwidth}
         \centering
    \includegraphics[width=0.8\textwidth]{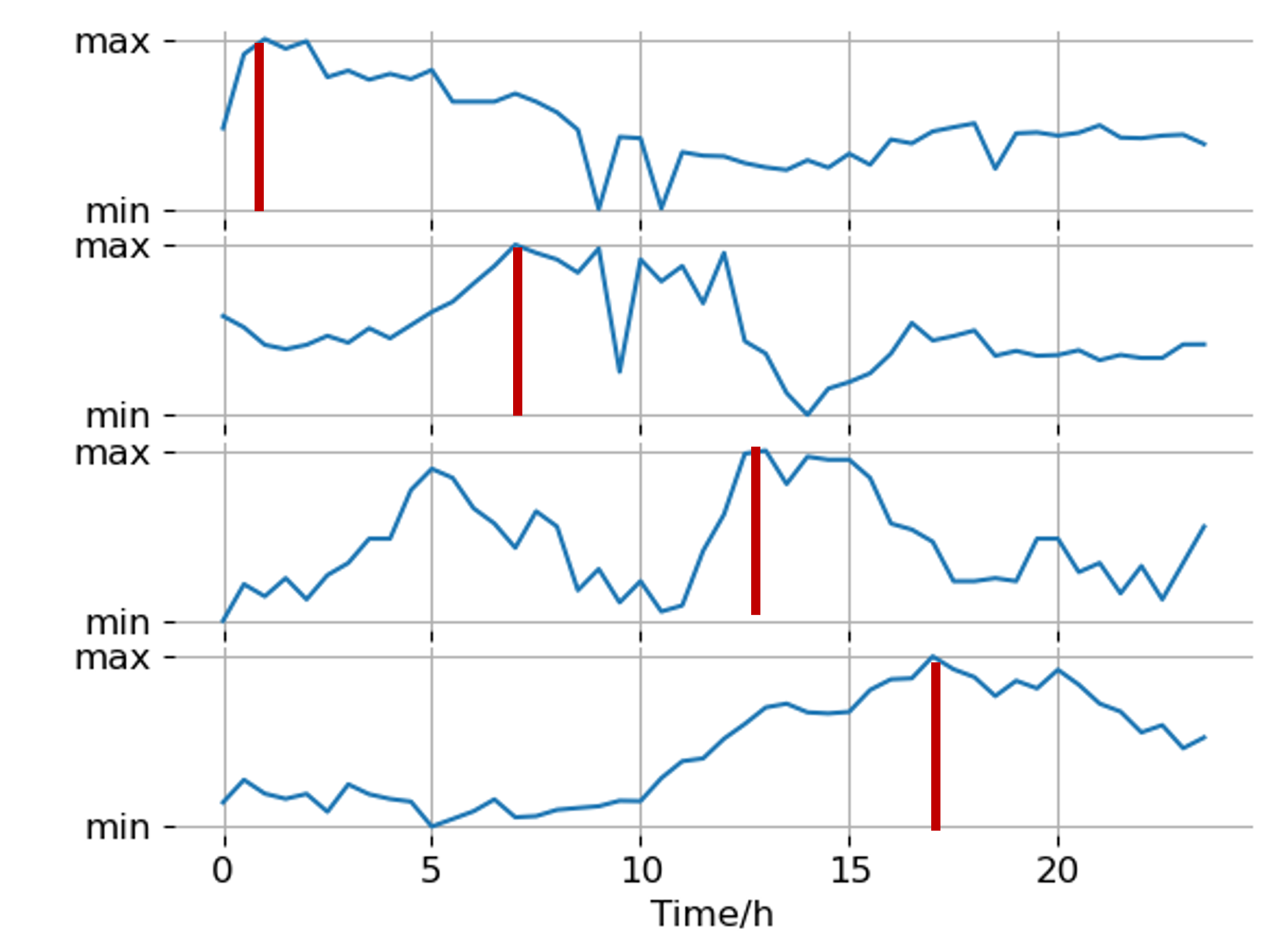}
         \caption{4 selected loads that peak at 1AM, 7AM, 1PM and 7PM, respectively.}
         \label{fig: load peaktime selected}
     \end{subfigure}
     \caption{Individual load patterns: spatial difference in peak load time.}
        \label{fig: individual load, peak time}
\end{figure}

\textbf{Observation 1:} Spatial differences exist in individual load variations. Figure \ref{fig: load peaktime dist} shows the distribution of peak time over approximately 10,000 loads. 
Results show that almost every time slot can witness some loads at their highest. Figure \ref{fig: load peaktime selected} further verifies this by showing four loads whose peak times occur at different times of the day.

\textbf{Physical reasons:} We attribute the spatial difference in load peak times to their different types and locations, as well as weather-related factors in different regions. Morning and evening can be peak hours for many residential and transportation loads, whereas depending on the type of consumers, some loads may use electricity during off-peak hours when the time-of-use rate is low and the price is cheaper.
%adopt time-of-use rate plans 

We further dive deeper into the difference in individual load variations.

\begin{figure}[h]
     \centering
\includegraphics[width=0.9\linewidth]{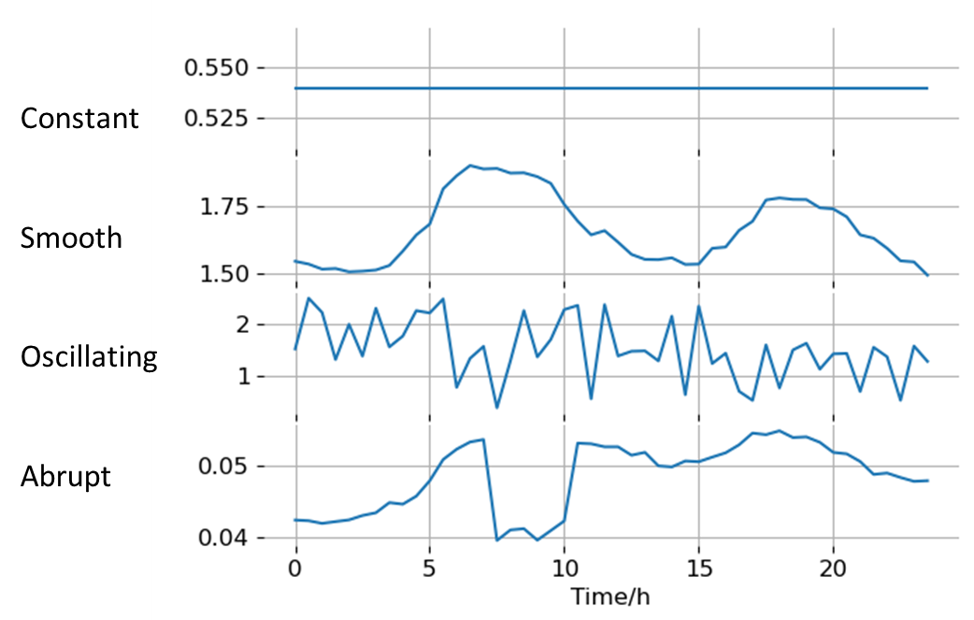}
     \caption{Individual load patterns: different styles of variation.}
        \label{fig: individual load style}
\end{figure}

\textbf{Observation 2:} Diverse styles of load variation exist. These styles include but are not limited to the following: constant load (load remains constant and stable throughout the day), smooth load (load changes smoothly), oscillating load (load changes frequently and noisily), and abrupt load (load changes abruptly at some point). Figure \ref{fig: individual load style} shows four loads of different styles.

\textbf{Physical reasons:} Differences can stem from the different types of loads in the power system. Residential loads consisting of household electrical appliances (lights, refrigerators, heaters, air conditioners, etc) may vary significantly in a day, with peak hours in the morning and evening (when people use more appliances for cooking, heating, air-conditioning, etc). Their off-peak hours are at night (when people sleep). Commercial loads like shopping and office can remain connected for longer durations of time, depending on their work schedule. Some industrial loads may have heavy machinery and systems that operate at all times, resulting in stable loads. Other types of loads can also be subject to unique patterns of variation. Some loads of traffic lighting work all times of day, making them nearly constant and stable. Loads of street lighting only operate at night, leading to abrupt changes at their startup/shutdown times, and stable consumption during their active period. Irrigation loads also work mostly during off-peak or night hours. And some transportation loads (like electric railways) can have peak hours in the morning and evening.

%see different electrical nature of loads: https://powersystemsinternational.com/types-of-electrical-loads/ 
% see  (primary energy consumption sectors: industrial, commercial, transportation, residential) https://www.energy.gov/eere/buildings/articles/commercial-miscellaneous-electric-loads-report-energy-consumption

Although a large abrupt change can sometimes occur on an individual load, it is still reasonable to assume that a load changes smoothly most of the time, based on our observations.

\textbf{Observation 3:} Figure \ref{fig: dPload dist, 0.5h} plots the distribution of individual load variations within a 0.5-hour interval. Results show that $90\%$ of the loads change by less than $8\%$ of their daily maximal value during any 0.5-hour interval. 

\begin{figure}[h]
\centering
\includegraphics[width=0.6\linewidth]{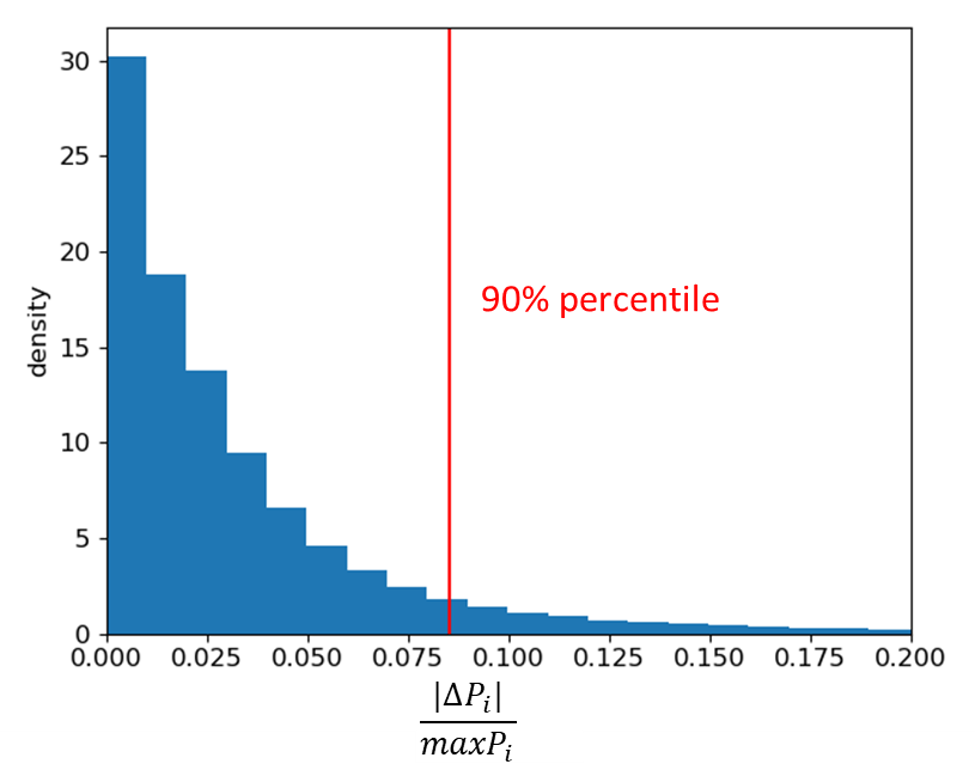}
\caption{Distribution of individual load variations within a 0.5-hour interval.}
\label{fig: dPload dist, 0.5h}    
\end{figure}

\subsection{Individual Generation Patterns}
\label{sec: patterns gen}

\begin{figure}[h]
\centering
\includegraphics[width=0.8\linewidth]{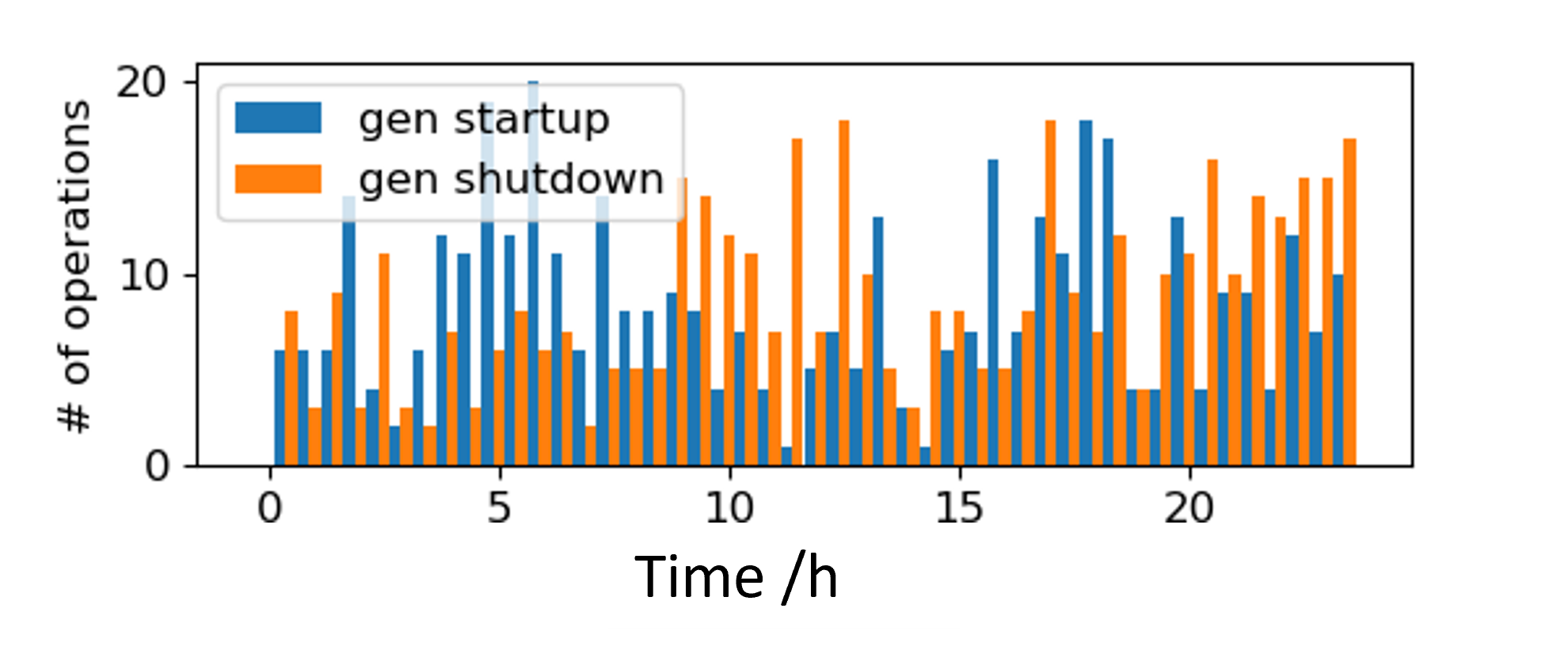}
\caption{Records of generation startups and shutdowns: a total of $>$300 operations observed in 24 hours, among approximately 2,000 active generation buses in total}
\label{fig: gen switching}    
\end{figure}

This section investigates individual generation patterns. 

\textbf{Observation:} The on/off switching of generation units can occur frequently throughout the day. In Figure \ref{fig: gen switching}, $>$300 status changes are observed in a day on a grid with approximately 2,000 generation buses.

\textbf{Physical reasons:} The change in generation status can be caused by either a control action or an unexpected and unplanned loss of generation (contingency). The control signal can be induced by a planned generation shutdown (for maintenance), and unit commitment (UC) operations performed hourly or sub-hourly\cite{UC-subhourly} in today's real-time market.

\begin{figure}[h]
\centering
\includegraphics[width=0.9\linewidth]{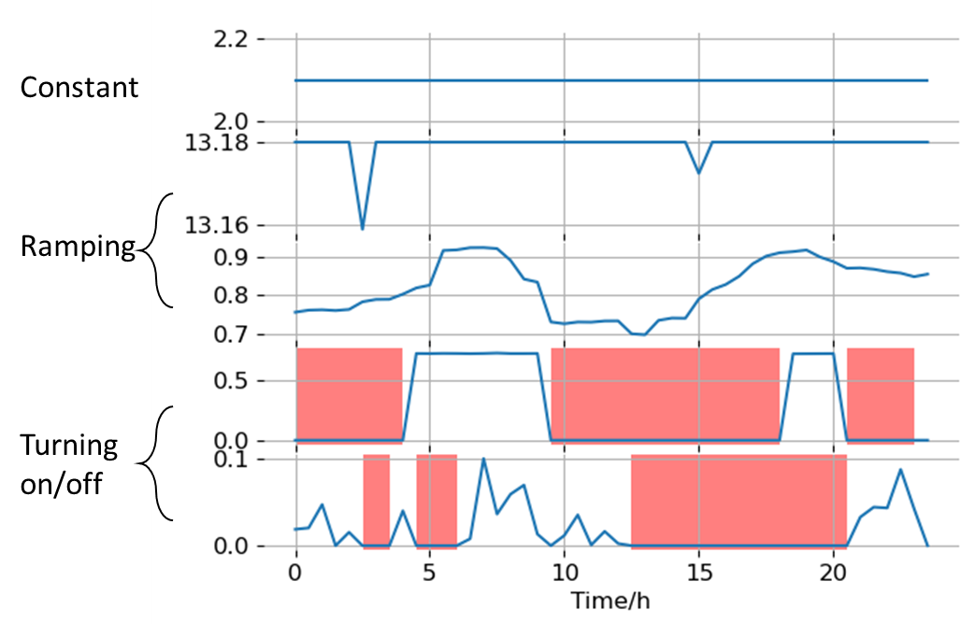}
\caption{Individual generation styles (red shades mark the periods when generation is off).}
\label{fig: individual gen style}    
\end{figure}

\textbf{Observation 2:} Different styles of generation variation also exist. Examples include constant generation (generation remains constant throughout the day), ramping generation (generation remains active but ramps up and down in a day due to power grid control), and generation with unit commitments (generation is turned on and off by the UC operations, and it can also ramp up and down). Figure \ref{fig: individual gen style} plots a few examples.

\textbf{Physical reasons:} 
The change in generation output can be caused by either: i) the nature of the energy resource, for instance, renewable energy resources have higher uncertainty, while conventional nuclear power plants are usually more stable, or ii) the automatic generation control (AGC) actions on synchronous generators induced by ED, OPF and the power system's frequency response.
Generators' status changes, if not a contingency, as discussed earlier, is usually decided by the ED and OPF control signals based on their prices (generation cost functions) and technical requirements.

\section{Potential Risks of Generalization}\label{sec:experiments}
\label{sec:Results}
%Data-driven methods can suffer from insufficient consideration of domain knowledge. This can significantly limit their practicality.
Ignoring the grid-specific topology, load, and generation patterns can cause poor generalization performance in data-driven models such that the outputs are inaccurate and even meaningless on unseen data. Specifically, we investigate the following generalization issues:
\begin{itemize}
    \item ML models ignoring system topology cannot generalize to dynamic graphs %(i.e., realistic behavior patterns of changing topology)
    \item ML models trained without consideration of spatial differences in load/generation cannot generalize well to realistic system configurations
\end{itemize}

\subsection{Risks of Generalization to Dynamic Graph}\label{sec: AD risk topology}
Many ML methods are based on non-graphical models. 
They cannot consider network topology and thus can result in bad estimates on real grids whose topology changes over time.
In this section, we evaluate the potential risks of generalization on dynamic graphs, from the perspective of data-driven time series anomaly detection. 

%Numerous univariate methods (hypothesis testings, auto-regressive and moving average models, etc.)  exist for anomaly detection in time series~\cite{keogh2007finding}. These methods apply to multivariate data when used together with dimension reduction techniques (principal component analysis, independent component analysis, etc.). Other methods have a multivariate nature and can learn spatial correlations in data.  Local outlier factor (LOF)~\cite{breunig2000lof} uses a local density approach. Isolation Forests~\cite{liu2008isolation}, as an emsemble method, partition the data using a set of trees for anomaly detection. Support vector machine (SVM) uses a classification method to make direct anomaly decisions.
%More recently, many approaches use neural networks~\cite{yi2017grouped}, distance-based~\cite{ramaswamy2000efficient}, and exemplars~\cite{jones2014anomaly}. %However, they hardly consider the graph structure. 

We evaluate a variety of outlier detection methods applied to time series grid data. For practical use, we select methods that are 1) \textit{unsupervised} (anomaly data are unlabeled in reality), and 2) \textit{online detection}: making decisions upon the arrival of new observations instead of requiring the future data to estimate for the present.
The experiment settings are in Appendix \ref{apdx: experiment setting, AD}. 
Below shows the different families of ML methods we evaluate:

\begin{table*}[]
\caption{Risks of generalization for anomaly detection.}
\label{tab: AD risk results}
\begin{tabular}{l|l|l}
\hline
\multirow{2}{*}{}                                                         & \multicolumn{1}{c|}{\textbf{Dynamic graph}} & \multicolumn{1}{c}{\textbf{\begin{tabular}[c]{@{}c@{}}Dynamic graph with \\ large spatial load differences\end{tabular}}} \\ \cline{2-3} 
    &
    \begin{subfigure}{0.4\textwidth}\centering\includegraphics[width=\textwidth]{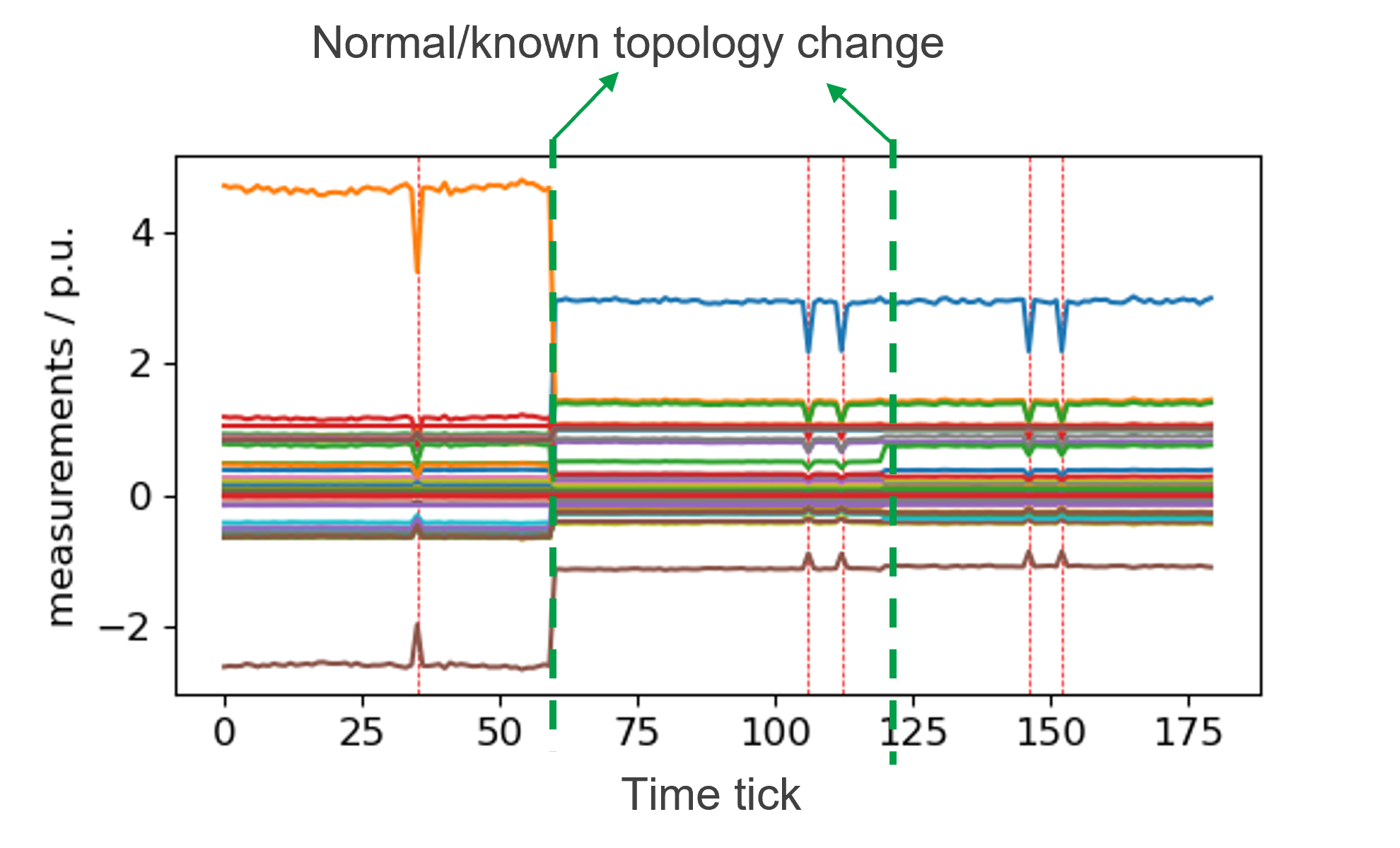}\caption{Time series measurement data drawn from dynamic graphs. The red vertical lines mark the anomalies, and the green lines mark the normal/known topology change}\label{fig: AD risk topology data}\end{subfigure}                                 
    & 
    \begin{subfigure}{0.4\textwidth}\centering\includegraphics[width=\textwidth]{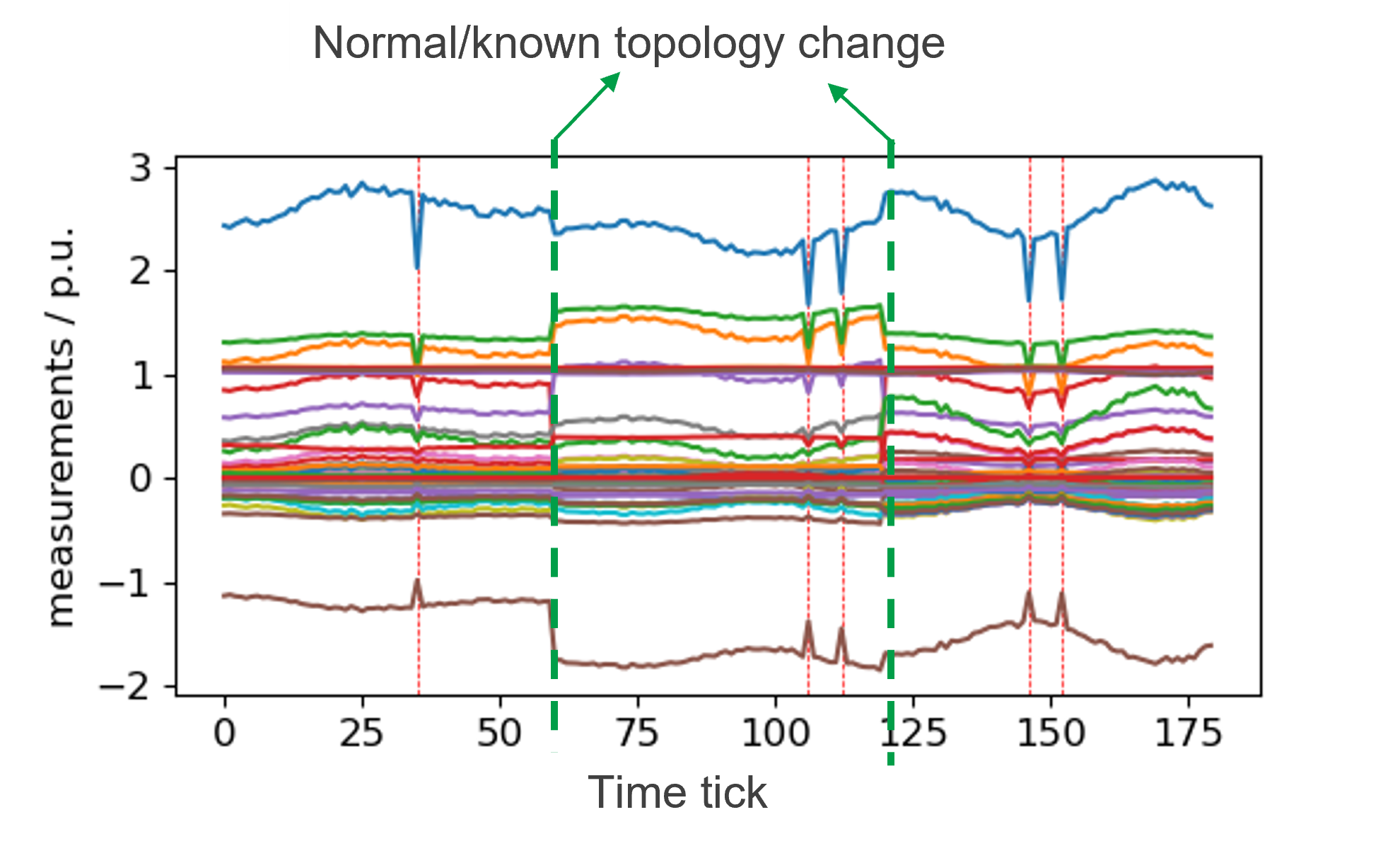}\caption{Time series measurement data drawn from dynamic graphs with large spatial differences and variations in load.}\label{fig: AD risk topology+realload data}\end{subfigure}  
    \\ \hline
\textbf{\begin{tabular}[c]{@{}l@{}}hypothesis test\\
%(online \\implementation)
\end{tabular}}        & \begin{subfigure}{0.4\textwidth}\centering
\includegraphics[width=\textwidth]{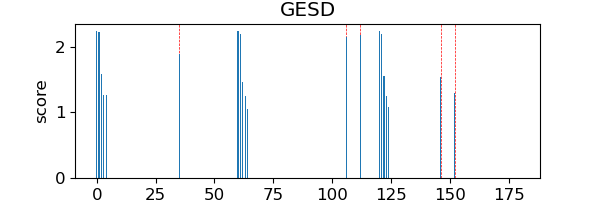}
\caption{GESD (sliding window) anomaly scores. False positive outcomes exist upon any normal topology change.}
\label{fig: AD risk gesd}  
\end{subfigure}                                &  \begin{subfigure}{0.4\textwidth}\centering
\includegraphics[width=\textwidth]{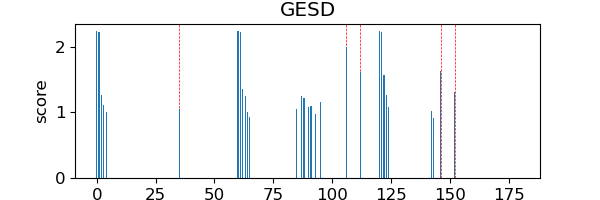}
\caption{More false positive outcomes under large spatial load differences.}
\label{fig: AD risk realload gesd}  
\end{subfigure}                                     \\ \hline
\textbf{\begin{tabular}[c]{@{}l@{}}autoregressive\end{tabular}}  
& \begin{subfigure}{0.4\textwidth}\centering
\centering
\includegraphics[width=\textwidth]{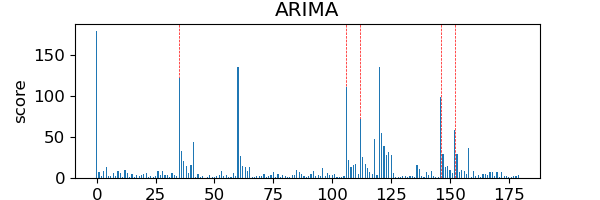}
     \hfill
\includegraphics[width=\textwidth]{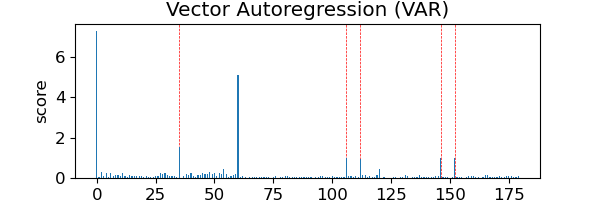}
     \caption{ARIMA and VAR anomaly scores. False positive outcomes appear upon any normal topology change.}
        \label{fig: AD risk autoreg}
\end{subfigure}                                            & 
\begin{subfigure}{0.4\textwidth}\centering
\centering
\includegraphics[width=\textwidth]{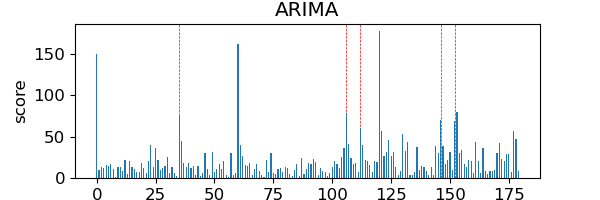}
     \hfill
\includegraphics[width=\textwidth]{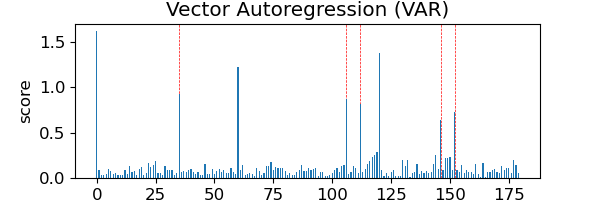}
     \caption{Models are harder to fit with large load variations. More false positives can occur.}
        \label{fig: AD risk realload autoreg}
\end{subfigure}  
\\ \hline
\textbf{\begin{tabular}[c]{@{}l@{}}density-based
\end{tabular}}  
&  
\begin{subfigure}{0.4\textwidth}\centering
\centering
\includegraphics[width=\textwidth]{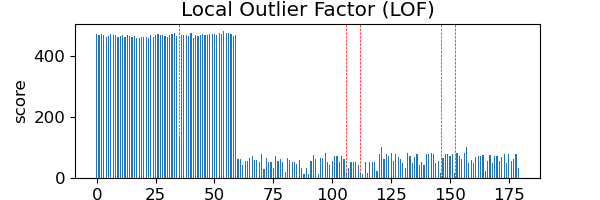}
\hfill     \includegraphics[width=\textwidth]{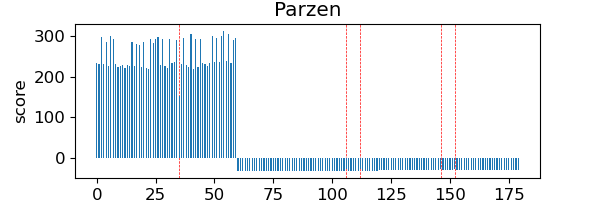}
     \caption{LOF and Parzen anomaly scores. Models fitted on some train data. Bad estimates exist on unseen topologies and make anomalies hardly detectable.}
        \label{fig: AD risk neighbor}  
\end{subfigure}
& 
\begin{subfigure}{0.4\textwidth}
\centering
\includegraphics[width=\textwidth]{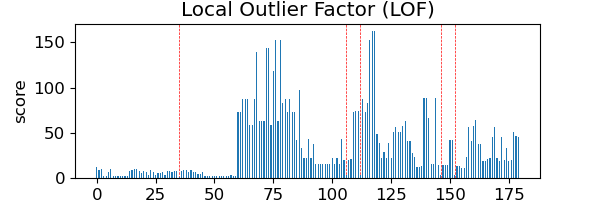}
\hfill     \includegraphics[width=\textwidth]{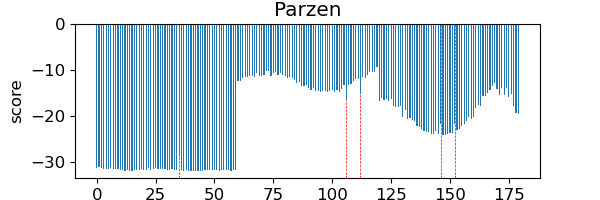}
     \caption{Bad and meaningless estimates appear on test data. The fitted model cannot generalize to unseen topology and new load configurations.}
        \label{fig: AD risk realload neighbor}  
\end{subfigure}                       \\ \hline
\end{tabular}
\end{table*}

\begin{table*}[]
\caption{Risks of generalization for anomaly detection. (Continued)}
\begin{tabular}{l|l|l}
\hline
\textbf{\begin{tabular}[c]{@{}l@{}}classification\end{tabular}}   
&                                      \begin{subfigure}{0.4\textwidth}
\centering
  \includegraphics[width=\textwidth]{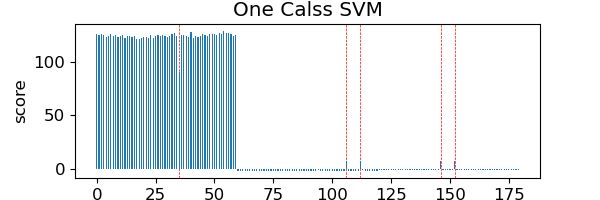}
     \caption{One-class SVM anomaly scores. Model fitted on train data. Bad estimates appear on the first unseen topology.}
        \label{fig: AD risk ocsvm}  
\end{subfigure}   
& \begin{subfigure}{0.4\textwidth}
\centering
  \includegraphics[width=\textwidth]{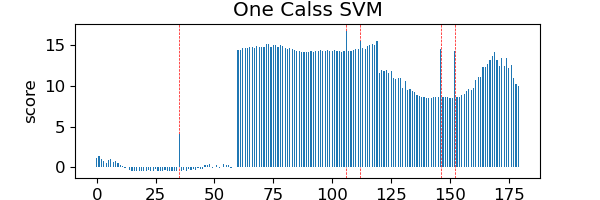}
     \caption{Bad estimates occur. The fitted model cannot generalize well to unseen topology and new load configurations.}
        \label{fig: AD risk realload ocsvm}  
\end{subfigure}                                     \\ \hline
\textbf{\begin{tabular}[c]{@{}l@{}}ensemble
\end{tabular}}        
&  \begin{subfigure}{0.4\textwidth}
\centering
\includegraphics[width=0.98\linewidth]{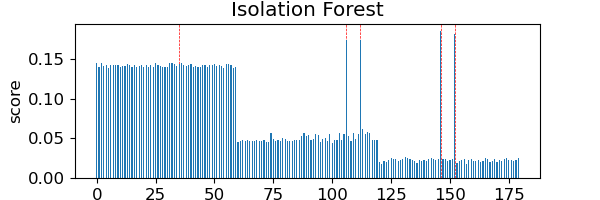}
\caption{Isolation Forest anomaly scores. Especially on the first
topology, the model gives high anomaly scores to all normal
data, making anomalies hardly detectable.}
\label{fig: AD risk iso}
\end{subfigure}
&  \begin{subfigure}{0.4\textwidth}
\centering
\includegraphics[width=0.98\linewidth]{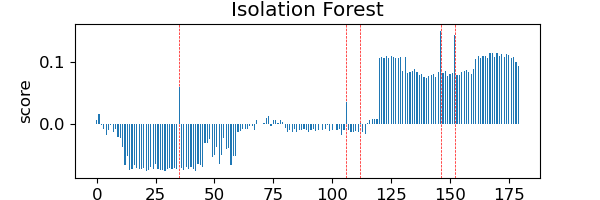}
\caption{Large anomalies scores on some new topology and load configurations. More false positives and some anomalies hardly detectable.}
\label{fig: AD risk realload iso}
\end{subfigure}                                     \\ \hline
\textbf{\begin{tabular}[c]{@{}l@{}}topology-aware \end{tabular}} & 
 \begin{subfigure}{0.4\textwidth}
\centering
\includegraphics[width=0.98\linewidth]{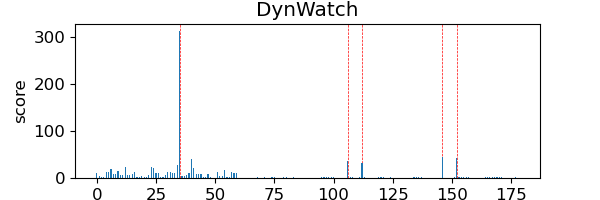}
\caption{DynWatch, a graph-distance-based time-series method is able to accurately detect anomalies on dynamic graphs without false alarms.}
\label{fig: AD risk dynwatch}
\end{subfigure}
&  \begin{subfigure}{0.4\textwidth}
\centering
\includegraphics[width=0.98\linewidth]{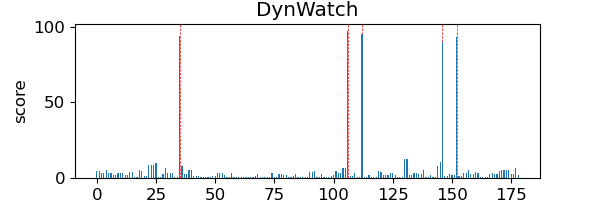}
\caption{DynWatch is able to tolerate load variations during anomaly detection. }
\label{fig: AD risk realload dynwatch}
\end{subfigure}                                                                                                                          \\ \hline
\end{tabular}
\end{table*}

% \begin{figure}[h]
% \centering
% \includegraphics[width=0.98\linewidth]{figures/ADrisks/measurements_marked.png}
% \caption{Time series measurement data for validation. The data are drawn from 3 different topologies. The red vertical lines mark the anomalies, and the green lines mark the normal/known topology change}
% \label{fig: AD risk topology data}    
% \end{figure}

\textbf{Hypothesis testing methods}: We evaluate the generalized extreme studentized deviation (GESD) test which is widely used in existing works for smart grid anomaly detection. The classic GESD is an offline algorithm that identifies the anomalous time ticks from the whole time series, based on R-statistics and critical values $\lambda$.
For online applications, we implement the method in sliding windows of width $50$, and the upper bound of the number of anomalies in a sliding window is assigned as $2$. 

%\textbf{Findings}: Figure \ref{fig: AD risk gesd} shows the anomalies scores on the unseen data. Result shows that, on a dynamic graph, the GESD creates false positive (FP) outcomes upon a normal topology change.

% \begin{figure}[h]
% \centering
% \includegraphics[width=0.98\linewidth]{figures/ADrisks/evalgesd_FDIA_.png}
% \caption{GESD (sliding window) anomaly scores. False positive outcomes exist upon any normal topology change.}
% \label{fig: AD risk gesd}    
% \end{figure}

\textbf{Auto-regressive (AR) models}: AR models forecast future data values based on a weighted sum of past observations. We evaluate a univariate method autoregressive integrated moving average (ARIMA) with hyperparameters $(p,d,q)=(5,1,0)$, and a multivariate method vector auto-regression (VAR) with $(max lags, differencing) = (5,1)$.

%\textbf{Findings}: Figure \ref{fig: AD risk autoreg} show the anomalies scores of ARIMA and VAR. Result shows that, on a dynamic graph, autoregressive models create false positive (FP) outcomes upon a normal topology change. 

% \begin{figure}[h]
%      \centering
%      \begin{subfigure}[t]{0.98\linewidth}
%          \centering
%     \includegraphics[width=\textwidth]{figures/ADrisks/arima_FDIA_.png}
%          \caption{}
%          \label{fig: AD risk arima}
%      \end{subfigure}
%      \hfill
%      \begin{subfigure}[t]{0.98\linewidth}
%          \centering
%     \includegraphics[width=\textwidth]{figures/ADrisks/var_FDIA_.png}
%          \caption{}
%          \label{fig: AD risk var}
%      \end{subfigure}
%      \caption{ARIMA and VAR anomaly scores. Auto-regressive models give false positive outcomes upon any normal topology change.}
%         \label{fig: AD risk autoreg}
% \end{figure}

\textbf{Density/neighborhood-based models}: We evaluate both the local outlier factor (LOF) and Parzen window estimator (also known as kernel density estimation) which are two well-known neighborhood-based anomaly detection methods.
LOF identifies anomalous data points that have a substantially lower density than their neighbors. The Parzen window estimator learns the probability density of every training data point using a kernel function.
In our experiments, LOF has $n_{neighbor}=3$ and the Parzen model uses a Gaussian kernel. To obtain good estimators in these approaches, the model needs to fit on a representative data set. So, we fit the models on our training set with 10 topologies and then evaluate them on the validation set which contains data from 3 unseen topologies.

%\textbf{Findings}: Figure \ref{fig: AD risk neighbor} show the anomalies scores of LOF and Parzen. Result shows that these models can create bad estimates on unseen topology. 
% \begin{figure}[h]
%      \centering
%      \begin{subfigure}[t]{0.98\linewidth}
%          \centering
%     \includegraphics[width=\textwidth]{figures/ADrisks/evallof_FDIA_.png}
%          \caption{}
%          \label{fig: AD risk lof}
%      \end{subfigure}
%      \hfill
%      \begin{subfigure}[t]{0.98\linewidth}
%          \centering
%     \includegraphics[width=\textwidth]{figures/ADrisks/evalparzen_FDIA_.png}
%          \caption{}
%          \label{fig: AD risk parzen}
%      \end{subfigure}
%      \caption{LOF and Parzen anomaly scores. Bad estimates on unseen topologies make anomalies hardly detectable.}
%         \label{fig: AD risk neighbor}
% \end{figure}

\textbf{Classification models}: We evaluate the one-class support vector machine (One-class SVM) model which is an unsupervised classifier to detect point outliers. It finds a decision boundary that separates high density of points (normal data) from those with small (anomalies). In our experiment, the model is trained with the RBF kernel and the upper bound on the fraction of training errors $\mu=0.5$. Then it is evaluated on the validation set.

%\textbf{Findings}: Figure \ref{fig: AD risk ocsvm} shows that One-class SVM creates bad estimates on certain unseen topology. 

% \begin{figure}[h]
% \centering
% \includegraphics[width=0.98\linewidth]{figures/ADrisks/evalocSVM_FDIA_.png}
% \caption{One-class SVM anomaly scores. Bad estimates on unseen topologies, especially on the first topology, and anomalies are undetectable.}
% \label{fig: AD risk ocsvm}    
% \end{figure}

\textbf{Ensemble models}: We implement an isolation forest which ensembles 100 trees, each drawing 1 feature from training data. And then, the model is evaluated on our validation set.

%\textbf{Findings}: Figure \ref{fig: AD risk iso} shows that isolation forest can create bad estimates on certain unseen topology. 

% \begin{figure}[h]
% \centering
% \includegraphics[width=0.98\linewidth]{figures/ADrisks/evaliso_FDIA_.png}
% \caption{Isolation Forest anomaly scores. Especially on the first topology, the model gives high anomaly scores to all normal data, making anomalies hardly detectable.}
% \label{fig: AD risk isolation}    
% \end{figure}

\textbf{Topology-aware methods}: We implement the DynWatch model proposed in \cite{dynwatch} which is an unsupervised time series online algorithm to detect point outliers on dynamic graphs. The model considers the impact of topology change using power sensitivity-based graph-distances. 

\textbf{Key findings}: Table \ref{tab: AD risk results} contains the anomaly scores produced by different families of methods. As results show that on a dynamic graph, the online implementation of the hypothesis test method GESD gives false positive (FP) outcomes. Univariate and multivariate auto-regression models suffer from similar false positive issues when the system shifts to a new topology. Density-based, classification-based, and isolation forests need to be fitted on a set of representative data before they can be used for inference. These methods do not model topology, and thus bad estimates appear when they are used on unseen data drawn from a different topology. And the more the new topology is different from the training data, the worse the estimates are. By contrast, the topology-aware method can detect anomalies  on dynamic graphs without generalization issues.

\textbf{Main conclusion}:  ML models ignoring topology cannot adapt to dynamic graphs. Potential risks for anomaly detection include producing false positive outcomes upon any normal topology change, as well as producing bad anomaly scores on unseen data from a new topology. For application on dynamic graphs, ML models need to be topology-aware.

\subsection{Risks of Generalization to Realistic Load}

Due to the limited availability of real data, many ML works are trained and evaluated on synthetic data. 
The realistic load behaviors have shown the spatial differences among individual loads, characterized by the different peak hours and diverse styles of variation. 
Such patterns in load variations are neglected in many works, leading to the generation of data in an unrealistic way.
As a result, a data-driven model trained and evaluated on unrealistic data can inevitably suffer from a lack of performance guarantees for practical use, raising the fear of making wrong decisions or infeasible predictions.
In this section, we evaluate the risk of generalization regarding the nonconsideration of spatial load differences, from the perspective of anomaly detection and data-driven AC-OPF. 

\subsection*{Risks on anomaly detection}
%As discussed in Section \ref{sec: patterns load}, an individual load have large variations and its own style making it differ significantly from others. 
In this Section, we evaluate how anomaly detection performance can be affected when large load variations and large spatial differences in load occur. Specifically, we compare the anomaly scores using voltage and current measurement data observed from a system with varying topology and different load patterns:

\begin{itemize}
    \item \textbf{Small load change and no spatial load difference}: At any time, all loads in the system are assumed to have strong correlations such that all loads ramp up and down at the same rate, without differences. The load variation is created from the real load traces in the public Lawrence Berkeley National Laboratory (LBNL) load data. Appendix \ref{apdx: experiment setting, AD} describes the data generation.
    \item \textbf{Load with large variation, spatial difference, and different styles}: This represents a more realistic individual load pattern. The system loads are divided into different groups, each containing a subset of loads with a certain style selected from the 4 observed styles in Figure \ref{fig: individual load style}.  Then load profiles are generated by interpolating the real-world individual load traces in the utility-provided data described in Section \ref{sec: grid conditions}. Figure \ref{fig: AD risk topology+realload data} shows the time-series data with realistic load patterns. The resulting data have large spatial differences and large load variations.
\end{itemize}

The experiment settings remain the same as Section \ref{sec: AD risk topology}

\textbf{Key findings}: The last column in Table \ref{tab: AD risk results} shows the anomaly detection performance of different ML models on dynamic graphs with spatial load differences. By comparison, we can see that when facing large variations and spatial differences in load, anomaly detectors  tend to create more false positive outputs. For models that require fitting on some train data, the fitted model cannot generalize well to unseen load configurations, creating bad anomaly scores that make anomalies less detectable.

\textbf{Main conclusion}: ML models for anomaly detection can face generalization risks under large load variations and spatial load differences, producing more false positive outcomes and bad anomaly scores on
unseen data. For better generalization, anomaly detection models need to build a tolerance to normal load variations.

\subsection*{Risks on data-driven ACOPF}
Considering the computational complexity of obtaining a large amount of labeled data for ACOPF, we evaluate a physics-constrained neural network\cite{dc3}\cite{homotopy-pnnl}. Unlike a supervised neural network (NN) which directly learns the input-output mapping from labeled data, the constrained neural network enables an unsupervised learning of solutions to nonlinear optimization problems, by enforcing network constraints on the output of NN. And the feasibility can be further enhanced by training with homotopy-based meta-optimization heuristics \cite{homotopy-pnnl}.

Now, suppose we have a series of $N_{real}$ realistic system configurations.
The realistic data comes from a power grid which can be divided into several sub-networks. Each sub-network contains a group of loads with the same individual load style (e.g., all loads are heavy industrial loads that correspond to a stable constant load style) and no spatial difference between them (all loads ramp up and down at the same rate). Whereas loads from different sub-networks have spatial differences, with different various styles and peak hours.
However, suppose that such a small dataset is insufficient, and we deploy data augmentation to create fake data.
Specifically, a subset of these realistic data is extracted as test samples, and the remaining data are augmented to create the training and validation set. The training and implementation of the neural network model are the same as that in \cite{homotopy-pnnl}. Now we compare 3 different ways of data generation (see detailed setting in Appendix \ref{apdx: experiment setting, opf}:
\begin{itemize}
    \item \textit{naive:} At any time moment, \textbf{all loads are positively correlated and  there is no spatial difference between individual load patterns.}
    \item \textit{grouped}: At any moment, loads are divided into groups.
    \textbf{Intra-group loads are correlated and inter-group loads have (random) spatial differences.}
    \item \textit{brute-force}: every individual load has its own pattern. \textbf{Random spatial differences exist everywhere. }
\end{itemize}
%\textbf{Interpretations}: These different ways of data augmentation represent different feature spaces chosen to draw synthetic data. Let the true feature space denote the feature space on which the true data distribution exists to generate all possible real-world cases. The given realistic data can be seen as some observations drawn from the true feature space. The naive method corresponds to a small subspace where a "special scenario" of positively correlated loads are fully exploited. This can cause non-representative data generated. The grouped method considers the intra-group load correlations and meanwhile  explores the unseen feature space via creating random inter-group spatial differences in load. This corresponds to a region that has a larger overlap with the true feature space. The brute-force method focuses on exploring a big feature space that is significantly larger than the true feature space. It enables considering any possible spatial load differences, however, inevitably creates a lot of samples outside the true feature space, and thus requires a huge amount of data for defining all possible realistic scenarios. 

Figure \ref{fig: acopf risk, case30} compares the risk of generalization in different data generation strategies, with different $N_{fake}$. 
The risk of generalization is quantified by the infeasibility of NN outputs, characterized by the mean violation of network constraints. We repeat the experiments 10 times with different random seeds $0,10,20,...,90$ in data generation in order to draw reliable conclusions.

\begin{figure}[h]
\centering
\includegraphics[width=0.8\linewidth]{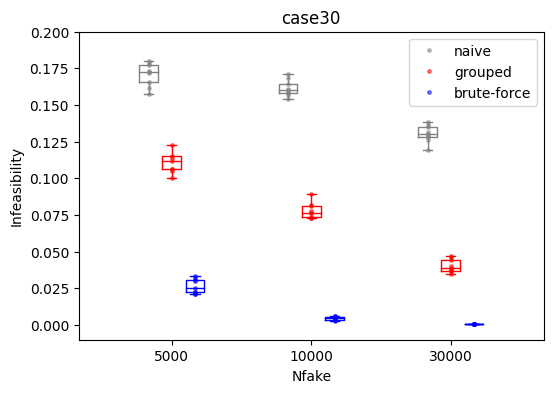}
\caption{Generalization risk of data-driven ACOPF on the 30-bus case (case30). $N_{real}=2000$. The NN has 2 layers and 200 hidden units in each.}
\label{fig: acopf risk, case30}    
\end{figure}
\begin{figure}[h]
\centering
\includegraphics[width=0.8\linewidth]{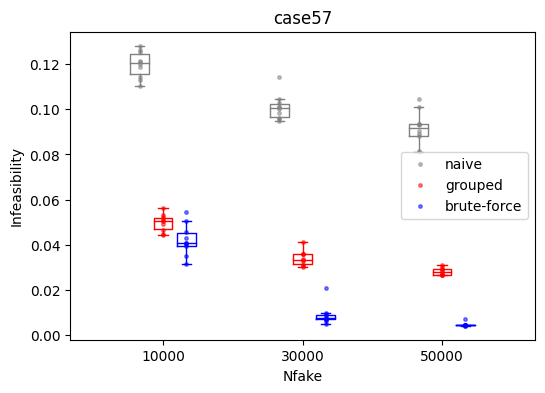}
\caption{Generalization risk of data-driven ACOPF on the 57-bus case (case57). $N_{real}=5000$. The NN has 4 layers and 200 hidden units in each.}
\label{fig: acopf risk, case57}    
\end{figure}

\textbf{Key findings:} Results in Figure \ref{fig: acopf risk, case30} and Figure \ref{fig: acopf risk, case57} show that the naive method which does not consider the spatial difference in data generation, results in models with poorest performance, giving highly infeasible outcomes; the grouped method has resulted in a lower risk of generalization; and the brute-force method which creates the most spatial difference in training data results in the lowest generalization risk on unseen data with realistic load patterns.

\textbf{Main conclusion}:  ML models trained and evaluated on unrealistic data ignoring individual load differences cannot generalize well on unseen system configurations with realistic load patterns. Such risk of generalization can result in highly infeasible predictions.

\section{Conclusion}
\label{sec:Conclusion}
%This paper encourages the incorporation of domain knowledge into data-driven methods through the analysis of real-world grid behaviors followed by the evaluation of the risks of machine learning generalization caused by insufficiently considering these realistic patterns.

This paper provides insights into the spatiotemporal grid patterns regarding the dynamic network topology as well as the temporal variations, spatial differences, and diverse styles in load and generation. 
Then, we evaluated the risks of generalization in data-driven models caused by ignoring these behavioral patterns. Key findings are that 1) ML models ignoring topology cannot adapt to dynamic graphs. This is verified by anomaly detectors giving false alarms on unseen topology, and 2) models trained on unrealistic load/generation patterns cannot generalize well on realistic power configurations. This is verified by anomaly detectors giving bad anomaly scores and data-driven OPF giving infeasible solutions when used on unseen data where large variations and spatial differences occur.
% \begin{itemize}
%     \item ML models ignoring topology cannot adapt to dynamic graphs. Our findings from time series anomaly detection show that ignoring topology can cause false positive outcomes upon any normal topology change as well as poor estimates on unseen data.
%     \item ML models trained and evaluated on data that are generated in an unrealistic way cannot generalize well to realistic system configurations.  Anomaly detection models tend to produce more false alarms under large load variations and spatial differences. Data-driven OPF model, when trained without consideration of spatial differences in load, tends to give highly infeasible outputs on realistic system configurations.
% \end{itemize}

 \begin{acks}

This research was supported in part by C3.ai Inc., Microsoft Corporation, and 
the Data Model Convergence (DMC) initiative
via the Laboratory Directed Research and Development (LDRD) investments at Pacific Northwest National Laboratory (PNNL). PNNL is a multi-program national laboratory
operated for the U.S. Department of Energy (DOE) by Battelle Memorial Institute under Contract
No. DE-AC05-76RL0-1830.

\end{acks}

\bibliographystyle{ACM-Reference-Format}
\bibliography{sample-base}


\begin{thebibliography}{20}

%%% the Web address.

\ifx \showCODEN    \undefined \def \showCODEN     #1{\unskip}     \fi
\ifx \showDOI      \undefined \def \showDOI       #1{#1}\fi
\ifx \showISBNx    \undefined \def \showISBNx     #1{\unskip}     \fi
\ifx \showISBNxiii \undefined \def \showISBNxiii  #1{\unskip}     \fi
\ifx \showISSN     \undefined \def \showISSN      #1{\unskip}     \fi
\ifx \showLCCN     \undefined \def \showLCCN      #1{\unskip}     \fi
\ifx \shownote     \undefined \def \shownote      #1{#1}          \fi
\ifx \showarticletitle \undefined \def \showarticletitle #1{#1}   \fi
\ifx \showURL      \undefined \def \showURL       {\relax}        \fi
% invisible to TeX
\providecommand\bibfield[2]{#2}
\providecommand\bibinfo[2]{#2}
\providecommand\natexlab[1]{#1}
\providecommand\showeprint[2][]{arXiv:#2}

\bibitem[Case(2016)]%
        {ukrain2015}
\bibfield{author}{\bibinfo{person}{Defense~Use Case}.}
  \bibinfo{year}{2016}\natexlab{}.
\newblock \showarticletitle{Analysis of the cyber attack on the Ukrainian power
  grid}.
\newblock \bibinfo{journal}{\emph{Electricity Information Sharing and Analysis
  Center (E-ISAC)}}  \bibinfo{volume}{388} (\bibinfo{year}{2016}).
\newblock


\bibitem[Donti et~al\mbox{.}(2021)]%
        {dc3}
\bibfield{author}{\bibinfo{person}{Priya~L Donti}, \bibinfo{person}{David
  Rolnick}, {and} \bibinfo{person}{J~Zico Kolter}.}
  \bibinfo{year}{2021}\natexlab{}.
\newblock \showarticletitle{DC3: A learning method for optimization with hard
  constraints}.
\newblock \bibinfo{journal}{\emph{arXiv preprint arXiv:2104.12225}}
  (\bibinfo{year}{2021}).
\newblock


\bibitem[Hooi et~al\mbox{.}(2018)]%
        {hooi2018gridwatch}
\bibfield{author}{\bibinfo{person}{Bryan Hooi}, \bibinfo{person}{Dhivya
  Eswaran}, \bibinfo{person}{Hyun~Ah Song}, \bibinfo{person}{Amritanshu
  Pandey}, \bibinfo{person}{Marko Jereminov}, \bibinfo{person}{Larry Pileggi},
  {and} \bibinfo{person}{Christos Faloutsos}.} \bibinfo{year}{2018}\natexlab{}.
\newblock \showarticletitle{GridWatch: Sensor Placement and Anomaly Detection
  in the Electrical Grid}. In \bibinfo{booktitle}{\emph{ECML-PKDD}}. Springer,
  \bibinfo{pages}{71--86}.
\newblock


\bibitem[Hu et~al\mbox{.}(2020)]%
        {pf-dnn-topo}
\bibfield{author}{\bibinfo{person}{Xinyue Hu}, \bibinfo{person}{Haoji Hu},
  \bibinfo{person}{Saurabh Verma}, {and} \bibinfo{person}{Zhi-Li Zhang}.}
  \bibinfo{year}{2020}\natexlab{}.
\newblock \showarticletitle{Physics-guided deep neural networks for power flow
  analysis}.
\newblock \bibinfo{journal}{\emph{IEEE Transactions on Power Systems}}
  \bibinfo{volume}{36}, \bibinfo{number}{3} (\bibinfo{year}{2020}),
  \bibinfo{pages}{2082--2092}.
\newblock


\bibitem[King et~al\mbox{.}(2022)]%
        {King2022}
\bibfield{author}{\bibinfo{person}{Ethan King}, \bibinfo{person}{Ján Drgoňa},
  \bibinfo{person}{Aaron Tuor}, \bibinfo{person}{Shrirang Abhyankar},
  \bibinfo{person}{Craig Bakker}, \bibinfo{person}{Arnab Bhattacharya}, {and}
  \bibinfo{person}{Draguna Vrabie}.} \bibinfo{year}{2022}\natexlab{}.
\newblock \showarticletitle{Koopman-based Differentiable Predictive Control for
  the Dynamics-Aware Economic Dispatch Problem}. In
  \bibinfo{booktitle}{\emph{2022 American Control Conference (ACC)}}.
  \bibinfo{pages}{2194--2201}.
\newblock
\urldef\tempurl%
\url{https://doi.org/10.23919/ACC53348.2022.9867379}
\showDOI{\tempurl}


\bibitem[Kotary et~al\mbox{.}(2021)]%
        {data-driven-optimization-survey}
\bibfield{author}{\bibinfo{person}{James Kotary}, \bibinfo{person}{Ferdinando
  Fioretto}, \bibinfo{person}{Pascal Van~Hentenryck}, {and}
  \bibinfo{person}{Bryan Wilder}.} \bibinfo{year}{2021}\natexlab{}.
\newblock \showarticletitle{End-to-end constrained optimization learning: A
  survey}.
\newblock \bibinfo{journal}{\emph{arXiv preprint arXiv:2103.16378}}
  (\bibinfo{year}{2021}).
\newblock


\bibitem[Kundacina et~al\mbox{.}(2022)]%
        {gnn-se}
\bibfield{author}{\bibinfo{person}{Ognjen Kundacina}, \bibinfo{person}{Mirsad
  Cosovic}, {and} \bibinfo{person}{Dejan Vukobratovic}.}
  \bibinfo{year}{2022}\natexlab{}.
\newblock \showarticletitle{State Estimation in Electric Power Systems
  Leveraging Graph Neural Networks}.
\newblock \bibinfo{journal}{\emph{arXiv preprint arXiv:2201.04056}}
  (\bibinfo{year}{2022}).
\newblock


\bibitem[Kërçi et~al\mbox{.}(2020)]%
        {UC-subhourly}
\bibfield{author}{\bibinfo{person}{T. Kërçi}, \bibinfo{person}{J. Giraldo},
  {and} \bibinfo{person}{F. Milano}.} \bibinfo{year}{2020}\natexlab{}.
\newblock \showarticletitle{Analysis of the impact of sub-hourly unit
  commitment on power system dynamics}.
\newblock \bibinfo{journal}{\emph{International Journal of Electrical Power \&
  Energy Systems}}  \bibinfo{volume}{119} (\bibinfo{year}{2020}),
  \bibinfo{pages}{105819}.
\newblock
\showISSN{0142-0615}
\urldef\tempurl%
\url{https://doi.org/10.1016/j.ijepes.2020.105819}
\showDOI{\tempurl}


\bibitem[Li et~al\mbox{.}({[n.\,d.]})]%
        {homotopy-pnnl}
\bibfield{author}{\bibinfo{person}{Shimiao Li}, \bibinfo{person}{Jan Drgona},
  \bibinfo{person}{Aaron~R Tuor}, \bibinfo{person}{Larry Pileggi}, {and}
  \bibinfo{person}{Draguna~L Vrabie}.} \bibinfo{year}{[n.\,d.]}\natexlab{}.
\newblock \showarticletitle{Homotopy Learning of Parametric Solutions to
  Constrained Optimization Problems}.
\newblock  (\bibinfo{year}{[n.\,d.]}).
\newblock


\bibitem[Li et~al\mbox{.}(2021)]%
        {dynwatch}
\bibfield{author}{\bibinfo{person}{Shimiao Li}, \bibinfo{person}{Amritanshu
  Pandey}, \bibinfo{person}{Bryan Hooi}, \bibinfo{person}{Christos Faloutsos},
  {and} \bibinfo{person}{Larry Pileggi}.} \bibinfo{year}{2021}\natexlab{}.
\newblock \showarticletitle{Dynamic graph-based anomaly detection in the
  electrical grid}.
\newblock \bibinfo{journal}{\emph{IEEE Transactions on Power Systems}}
  \bibinfo{volume}{37}, \bibinfo{number}{5} (\bibinfo{year}{2021}),
  \bibinfo{pages}{3408--3422}.
\newblock


\bibitem[Li et~al\mbox{.}(2022)]%
        {gridwarm_old}
\bibfield{author}{\bibinfo{person}{Shimiao Li}, \bibinfo{person}{Amritanshu
  Pandey}, {and} \bibinfo{person}{Larry Pileggi}.}
  \bibinfo{year}{2022}\natexlab{}.
\newblock \showarticletitle{GridWarm: Towards Practical Physics-Informed ML
  Design and Evaluation for Power Grid}.
\newblock \bibinfo{journal}{\emph{arXiv preprint arXiv:2205.03673}}
  (\bibinfo{year}{2022}).
\newblock


\bibitem[Li et~al\mbox{.}(2023)]%
        {gridwarm}
\bibfield{author}{\bibinfo{person}{Shimiao Li}, \bibinfo{person}{Amritanshu
  Pandey}, {and} \bibinfo{person}{Larry Pileggi}.}
  \bibinfo{year}{2023}\natexlab{}.
\newblock \bibinfo{title}{Contingency Analyses with Warm Starter using
  Probabilistic Graphical Model}.
\newblock
\newblock
\showeprint[arxiv]{2304.06727}~[cs.CR]


\bibitem[Liang et~al\mbox{.}(2016)]%
        {fdia-review}
\bibfield{author}{\bibinfo{person}{Gaoqi Liang}, \bibinfo{person}{Junhua Zhao},
  \bibinfo{person}{Fengji Luo}, \bibinfo{person}{Steven~R Weller}, {and}
  \bibinfo{person}{Zhao~Yang Dong}.} \bibinfo{year}{2016}\natexlab{}.
\newblock \showarticletitle{A review of false data injection attacks against
  modern power systems}.
\newblock \bibinfo{journal}{\emph{IEEE Transactions on Smart Grid}}
  \bibinfo{volume}{8}, \bibinfo{number}{4} (\bibinfo{year}{2016}),
  \bibinfo{pages}{1630--1638}.
\newblock


\bibitem[Mohammadian et~al\mbox{.}(2022)]%
        {physics-NN-pf}
\bibfield{author}{\bibinfo{person}{Mostafa Mohammadian}, \bibinfo{person}{Kyri
  Baker}, {and} \bibinfo{person}{Ferdinando Fioretto}.}
  \bibinfo{year}{2022}\natexlab{}.
\newblock \showarticletitle{Gradient-enhanced physics-informed neural networks
  for power systems operational support}.
\newblock \bibinfo{journal}{\emph{arXiv preprint arXiv:2206.10579}}
  (\bibinfo{year}{2022}).
\newblock


\bibitem[Ospina et~al\mbox{.}(2020)]%
        {MadIoTOnNY}
\bibfield{author}{\bibinfo{person}{Juan Ospina}, \bibinfo{person}{Xiaorui Liu},
  \bibinfo{person}{Charalambos Konstantinou}, {and} \bibinfo{person}{Yury
  Dvorkin}.} \bibinfo{year}{2020}\natexlab{}.
\newblock \showarticletitle{On the feasibility of load-changing attacks in
  power systems during the covid-19 pandemic}.
\newblock \bibinfo{journal}{\emph{IEEE Access}}  \bibinfo{volume}{9}
  (\bibinfo{year}{2020}), \bibinfo{pages}{2545--2563}.
\newblock


\bibitem[PJM(2022)]%
        {pjm-m03}
\bibfield{author}{\bibinfo{person}{PJM}.} \bibinfo{year}{2022}\natexlab{}.
\newblock \bibinfo{booktitle}{\emph{PJM Manual 03: Transmission Operations.
  Revision: 63}}.
\newblock
\urldef\tempurl%
\url{uments/manuals/m03.ashx}
\showURL{%
\tempurl}


\bibitem[PJM(2023)]%
        {pjm-m11}
\bibfield{author}{\bibinfo{person}{PJM}.} \bibinfo{year}{2023}\natexlab{}.
\newblock \bibinfo{booktitle}{\emph{PJM Manual 11: Energy and Ancillary
  Services Market Operations. Revision: 123}}.
\newblock
\urldef\tempurl%
\url{https://www.pjm.com/~/media/documents/manuals/m11.ashx}
\showURL{%
\tempurl}


\bibitem[Singer et~al\mbox{.}(2022)]%
        {grid-cybersecurity-Vyas}
\bibfield{author}{\bibinfo{person}{Brian Singer}, \bibinfo{person}{Amritanshu
  Pandey}, \bibinfo{person}{Shimiao Li}, \bibinfo{person}{Lujo Bauer},
  \bibinfo{person}{Craig Miller}, \bibinfo{person}{Lawrence Pileggi}, {and}
  \bibinfo{person}{Vyas Sekar}.} \bibinfo{year}{2022}\natexlab{}.
\newblock \showarticletitle{Shedding Light on Inconsistencies in Grid
  Cybersecurity: Disconnects and Recommendations}. In
  \bibinfo{booktitle}{\emph{2023 IEEE Symposium on Security and Privacy (SP)}}.
  IEEE Computer Society, \bibinfo{pages}{554--571}.
\newblock


\bibitem[Yadaiah and Sowmya(2006)]%
        {rnn-dse}
\bibfield{author}{\bibinfo{person}{Narri Yadaiah} {and} \bibinfo{person}{G
  Sowmya}.} \bibinfo{year}{2006}\natexlab{}.
\newblock \showarticletitle{Neural network based state estimation of dynamical
  systems}. In \bibinfo{booktitle}{\emph{The 2006 IEEE international joint
  conference on neural network proceedings}}. IEEE,
  \bibinfo{pages}{1042--1049}.
\newblock


\bibitem[Zamzam and Baker(2020)]%
        {dnn-opf-warm-starter}
\bibfield{author}{\bibinfo{person}{Ahmed~S Zamzam} {and} \bibinfo{person}{Kyri
  Baker}.} \bibinfo{year}{2020}\natexlab{}.
\newblock \showarticletitle{Learning optimal solutions for extremely fast AC
  optimal power flow}. In \bibinfo{booktitle}{\emph{2020 IEEE International
  Conference on Communications, Control, and Computing Technologies for Smart
  Grids (SmartGridComm)}}. IEEE, \bibinfo{pages}{1--6}.
\newblock


\end{thebibliography}

\appendix
\section{Appendix}
\subsection{Experiment settings for anomaly detection}\label{apdx: experiment setting, AD}

%\textbf{Experiment settings}:
In the experiments of anomaly detection, we create time series data with 13 different topologies, the first 10 of which are used as the training set for models that require a pre-fitting, and the last 3 of which are in the test set to evaluate the performance of all methods. 

The time series data are generated using the MATLAB tool. Specifically, the main procedures are:

\begin{enumerate}
    \item for each individual load on a test case, create time-series load data  from real-world load traces
    \item run power flow simulation using the MATPOWER toolbox based on the time-series load configurations
    \item simulate time-series sensor values of voltage, current, and power flow measurements by adding (Gaussian) noise on power flow solutions
\end{enumerate}
Specifically, the first step of creating synthetic time-series load data is necessary for a couple of reasons. In some cases, the real load data (like the real data described in Section \ref{sec: grid conditions}) are sampled with large time intervals (e.g., half-hourly intervals in the utility-provided case data in Section \ref{sec: grid conditions}), which are much larger than what we need in the experiment (we expect data sampled at second or less-than-one-minute intervals). Whereas in other cases, we might get load data at one location, or the total energy consumption of a power system, whereas we require time-series load profiles at each individual load locations on a certain power grid.

The main procedures to create a synthetic load trace of length $T_{new}$ time-ticks, from a real-world load trace of length $T_{ori}$ are as follows. More details are in \cite{dynwatch}.
\begin{enumerate}
    \item start from the original time-series load data of length $T_{ori}$(e.g., the 24-hour load profile at one particular load location, sampled at half-hourly interval)
    \item decompose the sequence into three parts: base $\mu$, variation $c(t)$, and noise $n(t)$ with $t=1,2,...,T_{ori}$, and learn a Gaussian noise distribution from the noise component
    \item apply interpolation on the variation component to generate a new  sequence $c_{new}(t)$ of length $T_{new}$: mathematically, $load_{new}=(1+\alpha c_{new}(t)+n_s(t))\mu$, noise $n_s(t)$ are sampled from the learned distribution and a scaling factor $\alpha$ is used to manipulate the magnitude of load variations
\end{enumerate}

In the experiments of Section \ref{sec: AD risk topology}, the dashed vertical lines mark the true anomaly moments, and green dashed lines marking the moments of known or expected topology changes. 

\subsection{Experiment settings for data-driven ACOPF}
\label{apdx: experiment setting, opf}:

The detailed settings of data generation in the experiments on data-driven ACOPF are as follows:
\begin{itemize}
    \item \textit{naive:} At one time moment, all loads in the system are scaled by the same randomly sampled load factor. I.e., for any new synthetic case $i$, randomly sample a load factor $c_i\in[0.8,1.2]$, a load $j$ is generated by $load(i,j) = c_i*\overline {load_j}, \forall j$, with  $\overline{load}_j$ denoting the base load $j$ obtained by taking the average on the realistic data. \textbf{This assumes all loads are positively correlated and results in no spatial difference between individual load patterns.}
    \item \textit{grouped}: loads are divided into groups (in the same way as the realistic data) and each group has a randomly sampled load factor to scale all loads it contains. I.e., for any new synthetic case $i$ and a load group $k$, we randomly sample $c_{ik}\in[0.8,1.2]$ and $load(i,j)=c_{ik}*\overline{load}_j, \forall j\in Group_k$
    \textbf{This results in intra-group spatial correlations and inter-group (random) spatial differences.}
    \item \textit{brute-force}: every individual load has its own pattern. I.e., for any new case $i$ and a load $j$, randomly sample $c_{ij}\in[0.8,1.2]$, and $load_(i,j) = c_{ij}*\overline{load}_j, \forall j$. \textbf{This creates random spatial differences everywhere. }
\end{itemize}

\end{document}